\documentclass{article}
\usepackage{package}

\usepackage{times}
\usepackage{soul}
\usepackage{url}
\usepackage[hidelinks]{hyperref}
\usepackage[utf8]{inputenc}
\usepackage[small]{caption}
\usepackage{graphicx}
\usepackage{amsmath}
\usepackage{amsthm}
\usepackage{booktabs}
\usepackage{algorithm}
\usepackage{algorithmic}
\usepackage[switch]{lineno}

\usepackage{enumitem}
\usepackage{multicol}
\usepackage{multirow}
\usepackage{subfig}
\usepackage{amssymb}
\usepackage{color}
\usepackage{xcolor}
\usepackage{balance}

\newtheorem{definition}{Definition}
\newtheorem{proposition}{Proposition}
\newtheorem{lemma}{Lemma}

\def\eqref#1{equation~\ref{#1}}

\def\1{\bm{1}}

\def\rmH{{\mathbf{H}}}

\def\rmX{{\mathbf{X}}}

\def\rmZ{{\mathbf{Z}}}

\DeclareMathAlphabet{\mathsfit}{\encodingdefault}{\sfdefault}{m}{sl}
\SetMathAlphabet{\mathsfit}{bold}{\encodingdefault}{\sfdefault}{bx}{n}

\def\gE{{\mathcal{E}}}

\def\gG{{\mathcal{G}}}
\def\gH{{\mathcal{H}}}

\def\gN{{\mathcal{N}}}

\def\gU{{\mathcal{U}}}
\def\gV{{\mathcal{V}}}
\def\gW{{\mathcal{W}}}

\def\sR{{\mathbb{R}}}

\newcommand{\R}{\mathbb{R}}

\newtheorem{theorem}{Theorem}

\title{Adaptive Expansion for Hypergraph Learning}

\author{Tianyi Ma$^1$ \and Yiyue Qian$^{2*}$ \and Shinan Zhang$^{2*}$ \and Chuxu Zhang$^3$ \And Yanfang Ye$^1$\\
\affiliations
$^1$ University of Notre Dame, Indiana, USA \\
$^2$ Amazon, Washington, USA\\
$^3$ University of Connecticut,  CT, USA \\
\emails
tma2@nd.edu, \{yyqian5,  zhangshinan\}@gmail.com,\\ chuxu.zhang@uconn.edu, yye7@nd.edu
}

\begin{document}

\maketitle
\footnotetext{The work is not related to the position at Amazon.}
\begin{abstract}
Hypergraph, with its powerful ability to capture higher-order relationships, has gained significant attention recently. 
Consequently, many hypergraph representation learning methods have emerged to model the complex relationships among hypergraphs. 
In general, these methods leverage classic expansion methods to convert hypergraphs into weighted or bipartite graphs, and further employ message passing mechanisms to model the complex structures within hypergraphs. 
However, classical expansion methods are designed in straightforward manners with fixed edge weights, resulting in information loss or redundancy.
In light of this, we design a novel clique expansion-based \textbf{Ad}aptive \textbf{E}xpansion method called \textbf{AdE} to adaptively expand hypergraphs into weighted graphs that preserve the higher-order structure information. 
Specifically, we introduce a novel Global Simulation Network to select two representative nodes for adaptively symbolizing each hyperedge and connect the rest of the nodes within the same hyperedge to the corresponding selected nodes.
Afterward, we design a distance-aware kernel function, dynamically adjusting edge weights to ensure similar nodes within a hyperedge are connected with larger weights.
Extensive theoretical justifications and empirical experiments over seven benchmark hypergraph datasets demonstrate that AdE has excellent rationality, generalization, and effectiveness compared to classic expansion models. Our source code is available \href{https://anonymous.4open.science/r/AdE-CEE2/README.md}{here}.
\end{abstract}

\section{Introduction}
\begin{figure}
\vspace{-1mm}
\begin{center}
    \includegraphics[scale=0.45]{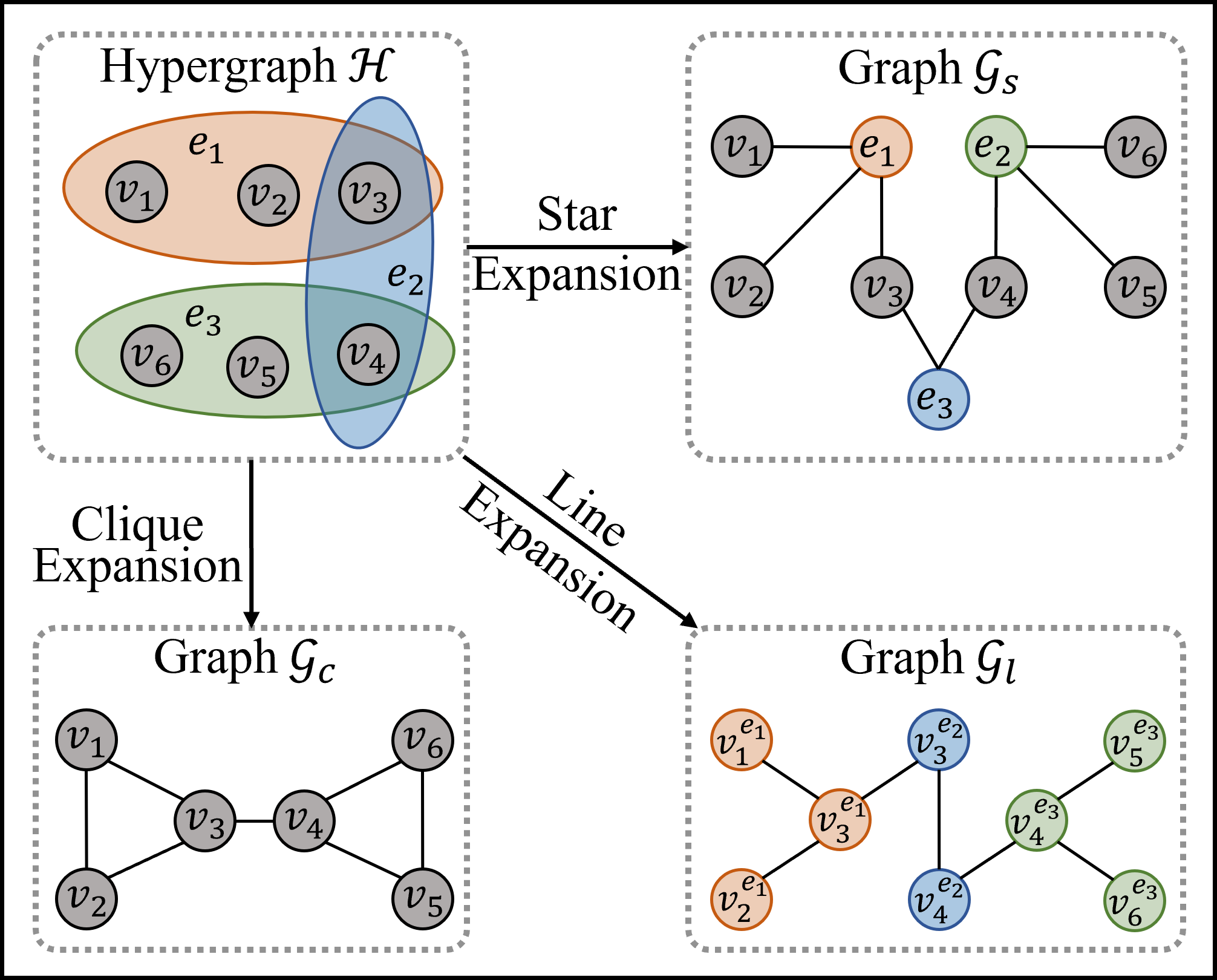}
\end{center}
\vspace{-3mm}
\caption{Illustration of classic hypergraph expansion methods.}
\label{fig: intro}
\vspace{-5mm}
\end{figure}

Hypergraphs, unlike pairwise relationships in graphs, introduce hyperedges to connect multiple nodes, enabling the representation of higher-order complex relationships. This concept has gained significant interest across multiple fields, such as social networks analysis, community detection, and recommendation systems~\cite{li2013link,zhang2022sparse,xia2022self,an2021hypergraph,ma2023hypergraph}. 
To leverage the benefits of hypergraphs, several hypergraph representation learning methods~\cite{DHNN,HRNN,Hyper-SAGNN,HyperGCL,liao2021hypergraph,qian2024dual,yan2024hypergraph} have been proposed to model the rich connectivity patterns within hypergraphs effectively. 
Generally speaking, most hypergraph representation learning methods primarily leverage classic expansion methods, e.g., clique expansion (CE)~\cite{CE}, star expansion (SE)~\cite{CESE}, and line expansion (LE)~\cite{LEGNN}, to convert hypergraphs into graphs, including bipartite graphs~\cite{Allset,xue2021multiplex,wu2022hypergraph} or weighted graphs~\cite{HyperGCN}, and employ message passing mechanisms to model the complex structures within hypergraphs.  
Specifically, CE substitutes hyperedges with cliques that edges connect every pair of nodes. 
For instance, HGNN~\cite{HGNN} utilizes a CE-based method, i.e., hypergraph Laplacian, to transfer the hypergraph into a weighted graph, and further employ a message-passing mechanism to learn the node representations.
The LE method~\cite{LEGNN} aims to construct a graph where nodes are made up of node-hyperedge pairs in the hypergraph, as illustrated in Fig.~\ref{fig: intro}. 
Two nodes are connected in the graph if they share the same node or hyperedge in the hypergraph. 

Although existing hypergraph representation learning with classic expansion methods achieves excellent performance in modeling complex relationships among nodes and hyperedges, these works~\cite{HyperGCN,HGNN,Hyper-SAGNN,LEGNN} still face the following limitations: (i) Some existing methods, such as CE-based methods, convert hypergraphs into graphs based on the hypergraph structure in a relatively straightforward manner, resulting in information loss or redundancy. 
For instance, the classic CE method connects every pair of nodes within the same hyperedge and makes it a fully connected subgraph, which brings the redundant information in the converted graph~\cite{CE}. 
(ii) Most methods employ the fixed edge weights when expanding hypergraphs into graphs while ignoring that nodes with similar attribute features are more likely to be connected with higher weights during the expansion. 
For example, HyperSAGE~\cite{HyperSAGE} converts hypergraphs into bipartite graphs where edge weights are uniformly assigned.
Afterward, HyperSAGE utilizes the hyperedge- and node-level message-passing strategy to propagate the information among nodes in the converted bipartite graphs. 
However, HyperSAGE ignores that nodes with similar attributes within the same hyperedge may have stronger connections during expansion.

To handle the aforementioned challenges, we propose a novel CE-based hypergraph expansion method called \textbf{Ad}aptive \textbf{E}xpansion (\textbf{AdE}) that expands hypergraphs into weighted graphs, depicting higher-order relationships among nodes. 
Specifically, to handle the first challenge, instead of
connecting all nodes within the same hyperedge indiscriminately, we design a novel Global Simulation Network (GSi-Net) to select two nodes for symbolizing each hyperedge adaptively. 
In particular, we first employ a pooling layer to obtain the global representations of the attribute features, followed by a simulation network to learn the importance of each feature dimension, and further obtain the adaptive weight matrix for the attribute features.  
After obtaining the scaled attribute feature with the adaptive weight matrix, we select two representative nodes for each hyperedge dynamically.
To address the second challenge, with the selected nodes for each hyperedge, we design a novel distance-aware kernel function to learn the edge weights to ensure that the edge between two nodes with similar attribute features will have a higher weight during hypergraph expansion. 
Last, the weighted graph can be fed to any graph neural networks (GNNs) for representation learning. 
To conclude, this work makes the following contributions:
\begin{itemize}[leftmargin=*]
    \item \textbf{\textit{Novelty:}} 
    We design a novel expansion method called AdE, including a GSi-Net and a distance-aware kernel function to expand hypergraphs into weighted graphs adaptively over different downstream tasks, which is the first work that learns to find the optimal graph structures adaptively during hypergraph expansion. 
    \item \textbf{\textit{Generalization:}} Our model is designed as a general expansion method that expands hypergraphs into weighted graphs, allowing powerful GNNs to effortlessly and seamlessly study hypergraphs. Theoretic justification demonstrates the generalization of AdE. 
    \item \textbf{\textit{Effectiveness:}} Theoretical justifications and empirical experiments over seven benchmark hypergraph datasets demonstrate the effectiveness of our model.
\end{itemize}

\section{Related Work}
\noindent\textbf{Hypergraph Expansions.} Unlike graphs where the edge connects two nodes, hypergraphs allow hyperedge to connect an arbitrary number of nodes which can depict high-order relationships. 
Most hypergraph representation learning methods first convert hypergraphs to graphs via classic expansion methods, i.e., clique expansion (CE)~\cite{CE}, star expansion (SE)~\cite{CESE}, 
line expansion (LE)~\cite{LEGNN} 
or its variants~\cite{HyperGCN,HGNN,zhou2006learning} and further feed the converted graph to neural networks for representation learning.
Clique expansion, one of the classic expansions, replaces hyperedges with cliques in which every node pair within the corresponding hyperedge is connected.
For instance, recent studies, i.e., HyperGCN~\cite{HyperGCN} and HGNN~\cite{HGNN} propose CE-based methods to convert hypergraphs into weighted graphs and learn the hypergraph representations for node classification tasks through graph neural networks (GNNs). 
Star expansion, another classic expansion, creates a set of nodes that represent hyperedges and further connects each new node with nodes that are in the corresponding hyperedge. 
Line expansion~\cite{LEGNN}, also called cross expansion~\cite{yan2024hypergraph}, constructs a graph with a new set of nodes where each node represents a node-hyperedge pair in the original hypergraph, and any two new nodes are connected if they share the same node or hyperedge on the hypergraph.
However, these methods fail to preserve high-order information within hypergraphs concisely and may lead to undesired losses in hypergraph conversion.  
Inspired by existing hypergraph expansions, this work proposes an adaptive hypergraph expansion to expand hypergraphs into weighted graphs to depict the higher-order relationships among nodes better. 

\noindent\textbf{Graph Neural Networks.}
 Graph neural networks (GNNs), considering both the node features and the graph structure, have become the state-of-the-art approach to learning graph representations~\cite{wang2025neuralgraphpatternmachine,wang2024learning,wang2024can}. 
 We would like to introduce several popular GNNs: GCN~\cite{GCN} implements a layer-wise propagation rule to learn the node embeddings from neighbors;
GAT~\cite{GAT} employs attention mechanisms to measure the importance of neighboring nodes when aggregating features;
GIN~\cite{GIN} is designed to utilize the Weisfeiler-Lehman Isomorphism Test~\cite{WLTest} for neighbor aggregation to enhance the capacity of GNNs in distinguishing different graph structures. 
Recently, more advanced GNNs~\cite{zhang2019heterogeneous,zhang2024ngqa,zhang2024mopi,GraphSage,wang2024gft,wang2024training} have been proposed, further demonstrating the effectiveness of GNNs.
 To leverage the powerful GNNs to learn the complex interaction in hypergraphs, inspired by existing works (e.g., HyperGCN and HGNN), we feed the weighted graph via our designed adaptive hypergraph expansion method to GNNs for downstream tasks.

\section{Preliminary}
\begin{definition}[Hypergraph]
Given a hypergraph $\mathcal{H}=(\mathcal{V}, \mathcal{E}, \mathcal{X})$, $\mathcal{V}$ is the set of nodes 
with size $N = |\mathcal{V}|$, $\mathcal{E}$ is the set of hyperedges with size $M = |\mathcal{E}|$, and $\mathcal{X}$ is the attribute feature set. 
Unlike the pairwise edge in a graph that only connects two vertices,  each hyperedge represents a higher-order interaction among a set of nodes. 
A hypergraph can be represented by an incidence matrix $\mathbf{H} \in \mathbb{R}^{N\times M}$, where $\mathbf{H}_{v,e} = 1$ if $ v\in e$; otherwise, $\mathbf{H}_{v,e} = 0$. 
Here, node $v \in \mathcal{V}$ and hyperedge $e \in \mathcal{E}$. Besides, we use $d(v) = \sum_{e\in\mathcal{E}} \mathbf{H}_{v,e}$ and $d(e) = \sum_{v\in\mathcal{V}} \mathbf{H}_{v,e}$ to denote the degrees of the node and hyperedge, respectively.
\end{definition}
\begin{definition}[Graph Neural Networks]~\label{def: gnn}
Most GNNs~\cite{wu2019simplifying,HGAN,RGCN,GTN,TailGNN} follow the neighbor aggregation operation in the messaging passing framework. Specifically, each node receives and aggregates the messages from the neighbor nodes recursively in multiple layers. 
For instance, the propagation rule of GNNs is formulated as follows:
\vspace{-1mm}
\begin{equation}
    \mathbf{Z}_{i,:}^{(l+1)} =  \mathcal{M}(\mathbf{Z}_{i,:}^l, \{\mathbf{Z}_{j,:}^l: v_j\in\mathcal{N}_{i}\}; W^{(l+1)} ), 
    \vspace{-1mm}
\end{equation}
where $\mathbf{Z}^{(l+1)}\in\R^{N\times d}$ is the $(l+1)$-th layer embeddings with $d$ dimensions, $\mathcal{N}_i$ is the neighbors of node $v_i$, and 
$\mathcal{M}(\cdot; W^{(l+1)})$ is the $(l+1)$-th message passing function with parameters $W$. 
Note that our framework is applicable to any GNN, e.g., GCN~\cite{GCN}.
\end{definition}

\noindent\textbf{Problem Definition.}
  \textit{
    Given a hypergraph $\mathcal{H}=(\mathcal{V}, \mathcal{E}, \mathcal{X})$, the objective is to design a hypergraph expansion method to convert the hypergraph to a weighted graph $\mathcal{G}_\text{a}$ and further employ GNNs with a mapping function $f_{\phi}:\mathcal{V}\rightarrow \mathbb{R}^{d}$ (with parameter $\phi$) to project each node $v_i \in \mathcal{V}$ to a $d$ dimensional embedding for downstream tasks with ground truth labels $Y$.}
    \vspace{-1mm}
\section{Methodology}
\begin{figure*}[t]
\centering
    \includegraphics[width=0.98\textwidth]{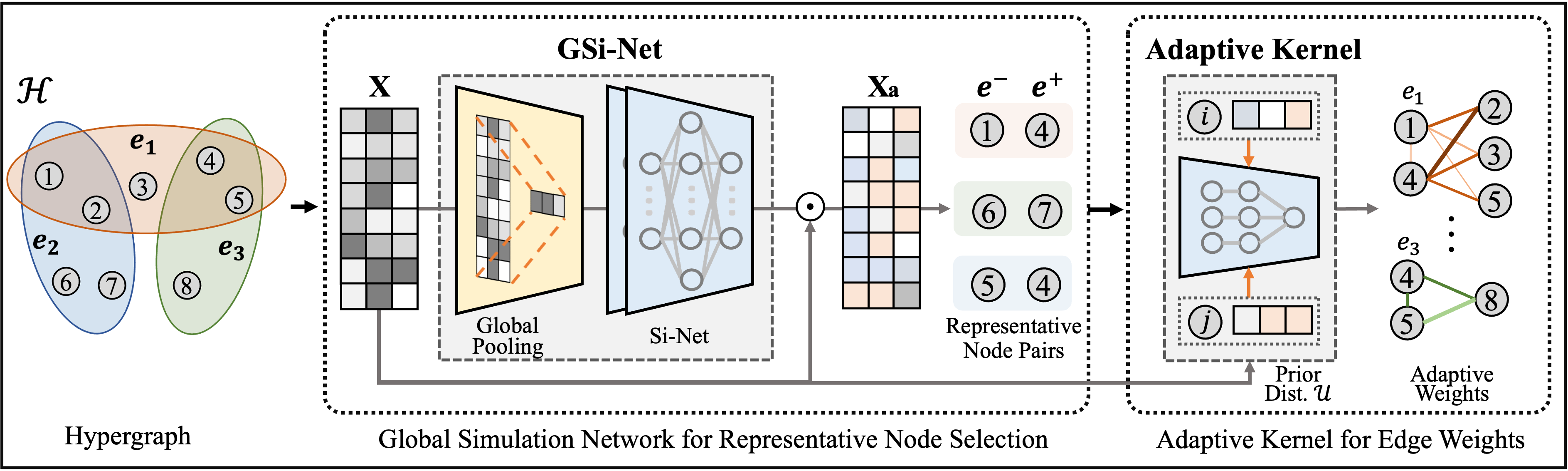}\label{fig: framework}
    \vspace{-2mm}
    \caption{The overall framework of AdE: given a hypergraph $\mathcal{H}=(\mathcal{V}, \mathcal{E}, \mathcal{X})$ with attribute feature $\mathbf{X}$, (i) AdE first feeds the attribute feature $\mathbf{X}$ into a global pooling layer, and leverages simulation network (Si-Net) to learn the importance of each feature dimension, further obtaining the weight matrix $W_g$. Afterward, AdE scales the attribute feature $\mathbf{X}_\text{a}$ with the learnable weight matrix $W_g$ and identifies a representative node pair $(v_{e^-}, v_{e^+})$ for each hyperedge $e$. 
    (ii) Using representative node pairs, AdE employs a distance-aware kernel function with prior distance information $\mathcal{U}$ to learn the edge weights adaptively and further constructs an adaptive weighted graph $\mathcal{G}_\text{a}$.}
    \vspace{-4mm}
\end{figure*}
In this section, we present the details of our \textbf{Ad}aptive \textbf{E}xpansion (\textbf{AdE}) to address the aforementioned limitations, which includes two key stages: (i) the \textbf{G}lobal \textbf{Si}mulation \textbf{Net}work (GSi-Net) for adaptive representative node selection and (ii) the distance-aware kernel function to learn the weights among node pairs in a dynamical manner. Moreover, we introduce extensive theoretical justifications to analyze the expressive power of our proposed method.

\subsection{GSi-Net for Adaptive Node Pair Selection} \label{section: gsi-net}

Previous works~\cite{HyperGCN,HGNN,HCHA,Allset,LEGNN} primarily convert hypergraphs into graphs via classic expansion methods, i.e., clique expansion (CE), star expansion (SE), line expansion (LE), and further leverage GNNs to learn the complex hypergraph structure. 
However, existing expansion methods convert hypergraphs into graphs based on the hypergraph structure in a straightforward manner, i.e., merely consider the hypergraph structures while ignoring the node attribute features in the expansion process. 
Moreover, the classic CE method is not concise enough and may lead to information redundancy as all pairs of nodes within the corresponding hyperedges will be connected. To handle this, HyperGCN proposes an updated CE-based expansion method that employs hypergraph Laplacian with mediators~\cite{chan2020generalizing} to expand hypergraph $\mathcal{H}$ to graph $\mathcal{G}$ by learning a nonlinear function over the real-valued signal $\mathbf{S}$~\cite{HyperGCN}. Specifically, HyperGCN first computes a real-valued signal $\mathbf{S}\in\mathbb{R}^N$, where $\mathbf{S}=\mathbf{X}\cdot \xi$. Here $\mathbf{X}\in\mathbb{R}^{N\times b}$ is the node attribute feature matrix with $b$ dimensions and $\xi\in \mathbb{R}^{b\times 1}$ is a random matrix.
With the signal $\mathbf{S}$, for each hyperedge $e$, HyperGCN selects two nodes $(v_{e^+}, v_{e^-})$ according to the mechanism: $(v_{e^+}, v_{e^-}) =\arg\max_{v_i,v_j\in e}|\mathbf{S}_i - \mathbf{S}_j|$. 
The node selection mechanism in HyperGCN is motivated by existing works~\cite{zhang2017re,zhou2006learning} that the pair of nodes with the longest distance on the hyperedge can represent the hyperedge information to a large extent. 
Then HyperGCN proposes to connect all other nodes in the hyperedge with the two representative nodes $v_{e^+}$ and $v_{e^-}$, respectively, forming an edge set $\gE_e$. 
Afterward, the weighted graph $\mathcal{G}$ is generated, where each edge is assigned a fixed weight $\frac{1}{2|e|-3}$ for further learning.

However, existing methods still face the following limitations in hypergraph expansion: (i) Due to the uncertainty of the random matrix $\xi$ for the real-value signal $\mathbf{S}$, the selected representative nodes for each hyperedge are biased, and not representative enough for the corresponding hyperedge. (ii) The edge weights among node pairs in $\mathcal{G}$ are fixed, which ignores the fact that two nodes with similar attribute features are more likely to be connected. To handle the first limitation, we design a \textbf{G}lobal \textbf{S}imulation \textbf{Net}work (GSi-Net) to select the most representative node pair for each hyperedge in an adaptive manner. For the second limitation, we design a distance-aware kernel function to dynamically adjust the edge weights in the weighted graph $\mathcal{G}_\text{a}$.

\noindent\textbf{Global Simulation Network.}
In consideration of the uncertainty of the random matrix for the real-value signal $\mathbf{S}$, we propose learning an informative attribute matrix that can be leveraged to dynamically identify the most characteristic node pair for hyperedges. 
Specifically, we first employ a mean pooling layer, to obtain the global-level representation of the attribute features $\mathbf{X}_{\text{g}} \in \mathbb{R}^{1\times b}$, where $\mathbf{X}_{\text{g}} = \frac{1}{N}\sum_{i=0}^{N} \mathbf{X}_{i,:}$ and $b$ is the feature dimension. 
Afterward, we design a simulation network (Si-Net) to adaptively learn the importance of each feature dimension, which is formulated as follows: 
\vspace{-1mm}
\begin{equation}
       W_{\text{g}} = \sigma(W_2\cdot\text{ReLU}(W_1 \cdot\mathbf{X}_{\text{g}})),
\end{equation}
where $W_{\text{g}} \in \mathbb{R}^{1\times b}$ is the learned importance matrix for all dimensions and $\sigma$ is the sigmoid activation function. 
Moreover, we scale the attribute feature $\mathbf{X}$ with the importance matrix $W_\text{g}$ and obtain the adaptive informative attribute matrix $\mathbf{X}_\text{a}$, where $\mathbf{X}_\text{a} = \mathbf{X}\odot W_\text{g}$. 
By dynamically learning the importance of each feature dimension, our adaptive attribute feature $\mathbf{X}_\text{a}$ is informative in selecting the representative node pair for the corresponding hyperedge.\\
\noindent\textbf{Representative Node Pair Selection.}
Instead of using the original attribute feature $\rmX$ with random noise $\xi$ as the real-value signal $\mathbf{S}$ in HyperGCN, we employ the sum of scaled attribute feature $\mathbf{X}_\text{a}$ among feature dimensions as the signal $\mathbf{S}=\sum_{k=1}^b \mathbf{X}_{\text{a},(:,k)}$, where $\mathbf{X}_{\text{a},(:,k)}$ is the $k$-th column of the scaled attribute feature $\mathbf{X}_\text{a}$ and $\mathbf{S}\in\mathbb{R}^{N}$, 
to adaptively select the most representative node pair to symbolize the corresponding hyperedge. 
In specific, for each hyperedge $e\in \mathcal{E}$, we select two nodes $(v_{e^-}, v_{e^+})$ to symbolize the hyperedge $e$ based on the following rule: $(v_{e^-}, v_{e^+}) = \arg\max_{v_i, v_j\in e} |\mathbf{S}_i-\mathbf{S}_j|$. 
Mention that, similar to existing works~\cite{chan2020generalizing,chan2018spectral,louis2015hypergraph}, if multiple node pairs satisfy the rule above, we randomly select one node pair as the representation of hyperedge $e$.
Subsequently, the remaining nodes within hyperedge $e$, denoted as $\mathcal{V}_m^e =\{v_m |v_m \ne v_{e^-}, v_m \ne v_{e^+}, v_m\in e\}$, are regarded as mediators. 
Moreover, each mediator in $\mathcal{V}_m^e$ will connect the representative nodes $v_{e^-}$ and $v_{e^+}$, respectively, and we further obtain the weighted graph $\mathcal{G}_\text{a}$ from hypergraph $\mathcal{H}$. 
The edge set in the weighted graph $\mathcal{G}_\text{a}$ for each hyperedge $e$ is denoted as $\mathcal{E}_e$, where 
$\mathcal{E}_e=\{\{v_{e^-}, v_{e^+}\},\{v_m, v_{e^-}\},\{v_m, v_{e^+}\}| v_m\in \mathcal{V}_m^e\}$. 
Unlike existing hypergraph expansion methods, i.e., CE, SE, and LE, that expand identical graph structures for all downstream tasks, our model AdE learns to generate the optimal graph by selecting the most representative node pairs adaptively for different downstream tasks.
\vspace{-1mm}
\subsection{Distance-aware Kernel for Edge Weight}
As mentioned previously, most existing methods assign the fixed edge weight to node pairs in the weighted graph~\cite{HyperGCN,HNHN,HyperSAGE}. 
However, it ignores the fact that node pairs with similar attribute features should have higher edge weights as they are more likely to be connected, compared with nodes with dissimilar attribute features. 
For instance, HyperGCN fixes the edge weight as $\frac{1}{2|e|-3}$, but does not consider that nodes within the same hyperedge may exhibit different behaviors.
Let's take an intuitive example of social media. 
Given a social media hypergraph, nodes are users and hyperedges describe a group of users who share the same interests. 
Assume a group of users is interested in photography and that two individuals focus on automotive photography while the rest are interested in landscape photography. In this case, when expanding the hypergraph to a social media graph, the fixed edge weight among these node pairs introduced by HyperGCN is inappropriate as all node pairs share the same edge weights. Intuitively, these two individuals having the same photography interest are supposed to get closer compared with the rest of the users within the group. 

In light of this, we design a distance-aware kernel function to learn the edge weight in an adaptive manner. 
Specifically, we first precompute the distance matrix $\mathcal{U}$  (i.e., Euclidean distance) among nodes based on the original attribute feature $\mathbf{X}$, where $\mathcal{U}_{i,j} = || \mathbf{X}_{i,:} - \mathbf{X}_{j,:} ||_2$. The prior information $\mathcal{U}$ is introduced to guide the learning process of edge weights. 
Afterward, we design a learnable kernel function to adjust the edge weight in $\mathcal{G}_\text{a}$, which is formulated as follows:
\begin{equation} \label{eq: weight func}
    \mathcal{W}_{i,j} = \exp(-\frac{1}{b}\sum\nolimits_{d=1}^b\frac{ \mathcal{U}_{i,j}(\mathbf{X}_{\text{a},({i,d})} - \mathbf{X}_{\text{a}, (j, d)})^2}{\theta_d^2}) ,
\end{equation}
where $\theta$ is a learnable matrix, $b$ is the dimension of $\mathbf{X}_\text{a}$, and $\mathbf{X}_{\text{a}, (i,d)}$ is the $d$-th element of the node $v_i$ in the adaptive attribute feature $\mathbf{X}_\text{a}$. Unlike existing hypergraph expansion methods that assign fixed weight to edges regardless of the downstream tasks, our novel distance-aware kernel function learns the edge weights dynamically. The edge weights between node pairs will be different if we handle different downstream tasks, which are realistic and practical in real-world scenarios.
Next, we would like to provide theoretical justification for our distance-aware kernel function. 
\begin{proposition} \label{proposition: w}
        The kernel function $\mathcal{W}$ learns to assign higher edge weights for node pairs with similar attribute features while smaller ones for less similar pairs.
\end{proposition}

\noindent We leave the justification of Proposition~\ref{proposition: w} in Appendix~\ref{proof: w}. Proposition~\ref{proposition: w} shows that our kernel function $\mathcal{W}$ is a distance-aware kernel function that adaptively assigns larger edge weights to similar node pairs and lower ones to less similar node pairs.
Inspired by the existing work~\cite{chan2020generalizing}, we standardize the edge weights such that the edge weights corresponding to hyperedge $e$ sum to 1. Formally, for each edge $\{v_i, v_j\} \in \mathcal{E}_e$, we compute the normalized weight $\Bar{\mathcal{W}}_{i,j}^{(e)}$ with respect to hyperedge $e$ as follows:
\vspace{-1mm}
\begin{equation} \vspace{-0.5mm} \label{eq: normalized weight function}
    \Bar{\mathcal{W}}_{i,j}^{(e)} = \frac{\mathcal{W}_{i,j}}{\sum_{\{v_k, v_g\}\in \mathcal{E}_e} \mathcal{W}_{k,g}}.
\end{equation}
Afterward, we obtain the adaptive weighted graph $\mathcal{G}_\text{a} = (\mathcal{V},\mathcal{E}_\text{a}, \mathcal{X}_\text{a})$ with the adaptive adjacency matrix $A_\text{a}$, where $A_{\text{a},(i,j)} = \sum_{e\in\mathcal{E}} \mathbb{I}\,[\{v_i, v_j\}\in\mathcal{E}_e]  \Bar{\mathcal{W}}_{i,j}^{(e)}$. 
\vspace{-1mm}
\subsection{Representation Learning} 
With the adaptive weighted graph $\mathcal{G}_\text{a} = (\mathcal{V}, \mathcal{E}_\text{a}, \mathcal{X}_\text{a})$, we employ powerful GNNs as graph encoders to learn the node embeddings over the adaptive weighted graph $\mathcal{G}_\text{a}$. Here, we take a two-layer GCN~\cite{GCN} as an example, denoted as AdE-GCN. The propagation rule of AdE-GCN to generate node embeddings is defined as follows:
\vspace{-1mm}
\begin{equation}~\label{eq: nodeembed}
    \mathbf{Z} = A_\text{a}\text{ReLU}(A_\text{a}\mathbf{X}_\text{a}W^{(1)})W^{(2)},
\end{equation}
where $A_\text{a}$ and $\mathbf{X}_\text{a}$ are the weighted adjacency matrix and adaptive attribute feature matrix in $\mathcal{G}_\text{a}$, respectively. 
Meanwhile, $W^{(1)}$ and $W^{(2)}$ are the learnable weight matrices for the first layer and second layer, respectively. Note that our model is designed as a general expansion method, which means that it can be effortlessly and seamlessly employed in any GNNs.\\ 
In this work, we employ the node classification task to evaluate the effectiveness of our designed method.
Thus, we leverage cross-entropy loss as our objective function to optimize parameters in AdE and the GNN encoder. 
The pseudocode of AdE is listed in the Appendix~\ref{appendix: algorithm}.
As our expansion design is motivated by HyperGCN, we provide propositions~\ref{proposition: 3} and~\ref{proposition: CE} to demonstrate that our expansion method expands a more effective graph than the existing method, i.e., HyperGCN.
We leave the justification of Proposition~\ref{proposition: 3} and~\ref{proposition: CE} in Appendix~\ref{proof: 3} and~\ref{proof: CE}, respectively.
\begin{proposition} \label{proposition: 3}
    Given the same selected nodes $(v_{e^-}, v_{e^+})$ for hyperedge $e$, our model AdE enhances HyperGCN by generating more adaptive weighted edges.
\end{proposition}
\begin{table*}[t]
\caption{Performance comparison (Mean accuracy ± std) of hypergraph expansion methods on GNNs for node classification. Bolded numbers indicate the best results of our models, and underlined numbers represent the best results of baseline methods.}
\vspace{-2mm}
\label{table: combine performance 1 3}
\centering
\setlength\tabcolsep{4pt}
\renewcommand{\arraystretch}{1.0}
\begin{tabular}{c|c|c|c|c|c|c|c|c|c|c}
\toprule
\multirow{2}{*}{Group} & \multirow{2}{*}{Model} & \multicolumn{5}{c|}{Homophilic Hypergraphs} & \multicolumn{4}{c}{Heterophilic Hypergraphs}  \\
\cmidrule{3-7} \cmidrule{8-11}
& & Cora-CA & DBLP & Cora & Citeseer & Pubmed & Senate 0.6 & Senate 1.0 & House 0.6 & House 1.0 \\
\midrule\midrule
\multirow{3}{*}{G1} & CE-GCN & 72.74\tiny{±1.19} & 86.26\tiny{±0.17} & 73.14\tiny{±1.09} & 67.23\tiny{±0.86} & 80.15\tiny{±1.02} & 50.94\tiny{±3.26} & 46.35\tiny{±3.34} & 68.57\tiny{±3.69} & 60.16\tiny{±3.40} \\
& CE-GAT & 73.85\tiny{±1.12} & 86.84\tiny{±0.12} & 74.00\tiny{±0.60} & 67.43\tiny{±0.77} & 81.26\tiny{±0.27} & 56.06\tiny{±1.78} & 53.59\tiny{±3.40} & 65.84\tiny{±4.27} & 63.20\tiny{±2.93} \\
& CE-GIN & 74.50\tiny{±1.24} & 87.89\tiny{±0.20} & 71.64\tiny{±0.60} & 68.36\tiny{±1.21} & 82.58\tiny{±0.36} & 54.71\tiny{±2.79} & 52.00\tiny{±5.37} & 67.26\tiny{±1.00} & 64.19\tiny{±1.60} \\
\midrule
\multirow{3}{*}{G2} & SE-GCN & 77.84\tiny{±1.00} & \underline{88.95\tiny{±0.18}} & 75.44\tiny{±0.83} & 68.25\tiny{±0.71} & 80.94\tiny{±0.42} & 47.82\tiny{±2.32} & 43.88\tiny{±2.78} & 62.74\tiny{±2.27} & 57.83\tiny{±2.93} \\
& SE-GAT & 78.61\tiny{±0.90} & 88.84\tiny{±0.21} & 75.33\tiny{±0.29} & \underline{68.65\tiny{±0.63}} & 81.41\tiny{±0.51} & 48.24\tiny{±3.65} & 45.47\tiny{±3.87} & 61.50\tiny{±5.36} & 56.82\tiny{±4.03} \\
& SE-GIN & 74.98\tiny{±0.82} & 87.18\tiny{±0.18} & 74.44\tiny{±0.38} & 68.15\tiny{±0.40} & 83.28\tiny{±0.62} & 51.29\tiny{±2.44} & 47.41\tiny{±2.02} & 67.08\tiny{±2.94} & 60.13\tiny{±1.99} \\
\midrule
\multirow{3}{*}{G3} & LE-GCN & 77.88\tiny{±1.77} & 88.77\tiny{±0.14} & 74.97\tiny{±0.71} & 66.78\tiny{±1.44} & 80.74\tiny{±0.34} & 50.35\tiny{±2.87} & 48.00\tiny{±2.58} & 64.65\tiny{±2.69} & 58.37\tiny{±3.22} \\
& LE-GAT & \underline{78.66\tiny{±0.57}} & 88.54\tiny{±0.24} & \underline{75.84\tiny{±0.66}} & 67.80\tiny{±0.55} & 80.95\tiny{±0.41} & 51.52\tiny{±3.21} & 45.26\tiny{±3.52} & \underline{68.96\tiny{±5.15}} & 62.82\tiny{±4.02} \\
& LE-GIN & 76.32\tiny{±1.36} & 88.31\tiny{±0.54} & 73.60\tiny{±0.61} & 67.73\tiny{±0.82} & 81.52\tiny{±0.58} & 53.82\tiny{±2.96} & 51.29\tiny{±2.35} & 66.40\tiny{±3.69} & 61.02\tiny{±3.52} \\
\midrule
\multirow{3}{*}{G4} & Uni-GCN & 75.21\tiny{±0.91} & 88.24\tiny{±1.04} & 75.54\tiny{±1.31} & 68.24\tiny{±0.35} & 83.51\tiny{±0.38} & 48.95\tiny{±3.82} & 45.02\tiny{±2.85} & 61.82\tiny{±3.71} & 57.25\tiny{±4.72} \\
& Uni-GAT & 76.53\tiny{±1.25} & 85.94\tiny{±1.47} & 74.37\tiny{±2.14} & 68.14\tiny{±0.57} & 83.20\tiny{±0.71} & \underline{59.69\tiny{±2.84}} & \underline{54.15\tiny{±5.44}} & 65.42\tiny{±2.95} & \underline{64.34\tiny{±2.46}} \\
& Uni-GIN & 77.05\tiny{±1.63} & 88.52\tiny{±1.06} & 75.24\tiny{±1.06} & 68.28\tiny{±1.02} & \underline{83.52\tiny{±0.62}} & 49.07\tiny{±3.83} & 45.94\tiny{±4.89} & 67.49\tiny{±5.35} & 60.59\tiny{±3.18} \\
\midrule\midrule
\multirow{3}{*}{Ours} & GCN & \textbf{81.42}\tiny{±1.16} & 89.68\tiny{±0.14} & \textbf{77.08}\tiny{±0.73} & \textbf{70.46}\tiny{±0.31} & 86.82\tiny{±0.64} & 49.07\tiny{±4.38} & 47.82\tiny{±3.54} & 62.48\tiny{±1.83} & 60.92\tiny{±2.58} \\
& GAT & 79.67\tiny{±0.80} & 90.24\tiny{±0.34} & 77.04\tiny{±0.55} & 70.12\tiny{±0.45} & 83.25\tiny{±0.41} & 56.71\tiny{±6.24} & 47.94\tiny{±3.74} & 78.84\tiny{±3.95} & 62.85\tiny{±3.02} \\
& GIN & 78.47\tiny{±0.99} & \textbf{90.54}\tiny{±0.21} & 74.44\tiny{±1.32} & 69.37\tiny{±0.55} & \textbf{86.94}\tiny{±0.43} & \textbf{68.54}\tiny{±3.17} & \textbf{59.82}\tiny{±1.72} & \textbf{81.09}\tiny{±2.05} & \textbf{66.49}\tiny{±2.62} \\
\bottomrule
\end{tabular}
\vspace{-2mm}
\end{table*}
\begin{proposition}\label{proposition: CE}
    Our model AdE is equivalent to the weight clique expansion in 3-uniform hypergraphs.
\end{proposition}
\subsection{Expressive Power of AdE}
We explore the expressive power of our proposed method AdE by building the bridge between the Weisfeiler-Leman algorithm (1-WL)~\cite{GIN,feng2024hypergraph,shervashidze2011weisfeiler} for the graph isomorphism test and the Generalized Weisfeiler-Leman Algorithm (1-GWL)~\cite{feng2024hypergraph,UniGNN} for the hypergraph isomorphism test~\cite{boker2019color}. 
\begin{definition}[Weisfeiler-Lehman test (1-WL test)]\label{def: 1-WL} 
Given a graph $\gG = (\gV_\gG, \gE_\gG)$, 1-WL initializes labels $l_i^{(0)}$ for every node $v_i\in \gV_\gG$. For each $t$-th iteration, the label of each node is updated as: $l_i^{(t+1)}=\text{HASH}\left(\{\{ (l_i^{(t)}, l_j^{(t)}) \}\}_{v_j\in\gN_i}\right)$. 
Here, $\text{HASH}(\cdot)$ is a function that assigns a new unique label to each unique set of labels, and $\gN_i$ is the neighbor set of node $v_i$. 1-WL test distinguishes two graphs $\gG=(\gV_\gG, \gE_\gG)$ and $\gG'=(\gV_{\gG'}, \gE_{\gG'})$ as non-isomorphic at time $t$ if $L(\gG) \ne L(\gG')$, where $L(G) = \{(l_i^{(t)}, count(v_i)) | v_i \in\gV_\gG\}$.
\end{definition}
\begin{definition}[Generalized Weisfeiler-Lehman test (1-GWL test)]\label{def: 1-GWL}
Given a hypergraph $\gH=(\gV_\gH,  \gE_\gH)$, 1-GWL assigns a unique label $h_i^{(0)}$ for each node $v_i\in\gV_\gH$ and an identical label $h_j^{(0)}$ for every hyperedge $e_j\in\gE_\gH$. 
Then, at $t$-th iteration, the labels are updated as: 
$h_e^{(t)}=\{\{ h_v^{(t)}\}\}_{v\in e}$, and $h_v^{(t+1)}=\{\{(h_v^{(t)}, h_e^{(t)})\}\}_{e\in \gE_{v}}$. 
Here, $\gE_{v}$ represents a set of hyperedges that contain node $v$. 
1-GWL test distinguish two hypergraphs $\gH=(\gV_\gH, \gE_\gH)$ and $\gH'=(\gV_{\gH'}, \gE_{\gH'})$ as non-isomorphic at iteration $t$, if $\{\{h_v^{(t)}|v\in \gV_\gH\}\}\ne \{\{h_u'^{(t)}|u\in \gV_{\gH'}\}\}$.
\end{definition}
Mention that, as 1-WL and 1-GWL tests ignore node features, we assign unified attribute features for all nodes in hypergraph $\gH$ so that AdE only depends on the hypergraph structure. Moreover, similar to this work~\cite{zhang2024expressive}, we are only concerned with node isomorphism classes. 
Hence, we remove the HASH function in 1-WL and merely consider the node labels. 
\vspace{-2mm}
\begin{lemma} \label{lemma: mapping}
         Given two hypergraph $\gH$ and $\gH'$, if AdE witnesses the same labels in 1-GWL for nodes $v_i\in\gV_\gH$ and $v_j\in\gV_{\gH'}$, i.e., $ h_{v_i}^{(t+1)}  = h_{v_j}^{(t+1)}$ at iteration $t > 0$, then it will generate the identical edges for nodes $v_i$ and $v_j$ in the converted graphs $\gG = AdE(\gH)$ and $\gG'=AdE(\gH')$, respectively.
\end{lemma}
\begin{theorem}\label{theorem 1GWL 1WL}
 Given two hypergraphs $\gH$ and $\gH'$, if the 1-GWL test cannot distinguish hypergraphs $\gH$ and $\gH'$ within $T$ iterations, then the 1-WL test is also unable to differentiate the converted graphs $\gG = \text{AdE}(\gH)$ and  $\gG' = \text{AdE}(\gH')$.
\end{theorem}
\begin{theorem}\label{theorem 1WL 1GWL}
  Given two hypergraphs $\gH$ and $\gH'$, if 1-WL test decides the converted graphs $\gG = \text{AdE}(\gH)$ and $\gG' = \text{AdE}(\gH')$ are non-isomorphic at iteration $t>0$, then 1-GWL test also decides $\gH\ne\gH'$.
\end{theorem}
\indent The theoretical justification of Lemma~\ref{lemma: mapping}, Theorem~\ref{theorem 1GWL 1WL}, and Theorem~\ref{theorem 1WL 1GWL} are provided in Appendix~\ref{proof: mapping}, ~\ref{proof: 1GWL 1WL}, and ~\ref{proof: 1WL 1GWL}, respectively.
\section{Experiments}
\begin{table*}[t]
\caption{Performance comparison (Mean accuracy \% ± std) of HyGNNs for node classification. Bolded numbers indicate the best results, and underlined numbers represent the runner-up performance.}
\vspace{-2mm}
\label{table: combine performance 2 4}
\centering
\setlength\tabcolsep{6pt}
\renewcommand{\arraystretch}{1.0}
{\small
\begin{tabular}{c|c|c|c|c|c|c|c|c|c}
\toprule
\multirow{2}{*}{Model} & \multicolumn{5}{c|}{Homophilic Hypergraphs} & \multicolumn{4}{c}{Heterophilic Hypergraphs} \\
\cmidrule{2-10}  &  Cora-CA & DBLP & Cora & Citeseer & Pubmed & Senate 0.6 & Senate 1.0 & House 0.6 & House 1.0 \\
\midrule\midrule
MLP & 69.83\tiny{$\pm$0.82} & 82.84\tiny{$\pm$0.19} & 68.45\tiny{$\pm$0.78} & 68.01\tiny{$\pm$1.86} & 74.45\tiny{$\pm$0.31} & 62.24\tiny{$\pm$6.39} & 50.35\tiny{$\pm$3.25} & 77.12\tiny{$\pm$3.00} & 64.07\tiny{$\pm$2.70} \\
HGNN & 79.73\tiny{$\pm$0.63} & 88.42\tiny{$\pm$0.31} & 76.11\tiny{$\pm$0.85} & 69.32\tiny{$\pm$0.50} & 81.53\tiny{$\pm$0.25} & 60.06\tiny{$\pm$2.81} & 49.35\tiny{$\pm$2.64} & 61.24\tiny{$\pm$1.72} & 57.24\tiny{$\pm$1.76} \\
HCHA & 79.62\tiny{$\pm$0.64} & 88.02\tiny{$\pm$0.27} & \underline{75.97\tiny{$\pm$0.90}} & 68.84\tiny{$\pm$1.12} & 81.34\tiny{$\pm$1.04} & 43.74\tiny{$\pm$1.52} & 46.20\tiny{$\pm$2.84} & 61.28\tiny{$\pm$1.54} & 56.98\tiny{$\pm$1.42} \\
HyperGCN & 75.64\tiny{$\pm$2.08} & 88.68\tiny{$\pm$0.29} & 73.89\tiny{$\pm$1.16} & 68.08\tiny{$\pm$1.17} & 77.70\tiny{$\pm$3.52} & 52.18\tiny{$\pm$2.75} & 51.82\tiny{$\pm$2.49} & 75.62\tiny{$\pm$2.03} & 62.43\tiny{$\pm$2.68} \\
HNHN & 72.32\tiny{$\pm$1.52} & 86.82\tiny{$\pm$0.14} & 71.52\tiny{$\pm$1.47} & 68.36\tiny{$\pm$1.24} & 79.63\tiny{$\pm$0.52} & 62.18\tiny{$\pm$6.99} & 54.71\tiny{$\pm$4.15} & 68.36\tiny{$\pm$2.21} & 65.16\tiny{$\pm$1.88} \\
AllSet & 78.52\tiny{$\pm$1.84} & 89.95\tiny{$\pm$0.12} & 74.26\tiny{$\pm$0.76} & 69.71\tiny{$\pm$1.00} & 85.63\tiny{$\pm$1.62} & 65.47\tiny{$\pm$3.42} & 50.82\tiny{$\pm$4.60} & 78.95\tiny{$\pm$1.46} & 65.20\tiny{$\pm$1.58} \\
ED-HNN & \underline{79.80\tiny{$\pm$0.59}} & \underline{89.98\tiny{$\pm$0.21}} & 75.78\tiny{$\pm$1.26} & \underline{69.82\tiny{$\pm$2.77}} & \underline{85.62\tiny{$\pm$1.32}} & \underline{65.53\tiny{$\pm$3.10}} & \underline{55.47\tiny{$\pm$4.87}} & \underline{79.01\tiny{$\pm$1.00}} & \underline{65.70\tiny{$\pm$1.98}} \\
SheafHGNN &  78.42\tiny{$\pm$1.94} & 89.76\tiny{$\pm$0.72}&73.52\tiny{$\pm$1.50}&69.77\tiny{$\pm$1.92}&83.42\tiny{$\pm$2.76}&64.15\tiny{$\pm$2.55}&51.25\tiny{$\pm$4.16}&77.72\tiny{$\pm$3.81}&64.94\tiny{$\pm$1.45}\\
DPHGNN & 77.96\tiny{$\pm$1.73}&89.45\tiny{$\pm$1.43}&74.37\tiny{$\pm$2.06}&68.28\tiny{$\pm$3.95}&82.37\tiny{$\pm$1.74}&63.17\tiny{$\pm$2.29}&53.65\tiny{$\pm$3.22}&78.82\tiny{$\pm$2.98}&64.36\tiny{$\pm$1.60}\\
\midrule\midrule
Ours & \textbf{81.42\tiny{$\pm$1.16}} & \textbf{90.54\tiny{$\pm$0.21}} & \textbf{77.08\tiny{$\pm$0.73}} & \textbf{70.46\tiny{$\pm$0.31}} & \textbf{86.82\tiny{$\pm$0.64}} & \textbf{68.54\tiny{$\pm$3.17}} & \textbf{59.82\tiny{$\pm$1.72}} & \textbf{81.09\tiny{$\pm$2.05}} & \textbf{66.49\tiny{$\pm$2.62}} \\
\bottomrule
\end{tabular}
}
\vspace{-4mm}
\end{table*}

In this section, we first introduce the experimental setup and then compare AdE with various baseline methods to show its effectiveness.
Moreover, ablation studies, embedding visualizations, and complexity analyses are conducted to show the rationality, effectiveness, and efficiency of AdE.
\vspace{-1mm}
\subsection{Experimental Setup}
To evaluate the effectiveness of our model, we employ five benchmark homophilic hypergraph datasets from~\cite{Allset}, including Cora-CA, DBLP, Cora, Citeseer, and Pubmed, and two heterophilic hypergraph datasets, House and Senate~\cite{chodrow2021hypergraph,wang2022equivariant}, with feature noise $0.6$ and $1.0$. 
A detailed discussion about datasets and data statistics is provided in Appendix~\ref{appendix: data}.

\noindent\textbf{Baseline Methods.} We compare our model with four classic hypergraph expansion methods, including clique expansion (CE) based methods (G1), star expansion (SE) based methods (G2), line expansion (LE) based methods (G3), and Uni-based methods (G4). 
We leverage three GNNs, i.e., GCN~\cite{GCN}, GAT~\cite{GAT}, and GIN~\cite{GIN}, as backbone models over the converted graphs. 
Besides, we also conduct experiments on MLP and eight HyGNNs models, including HGNN~\cite{HGNN}, HCHA~\cite{HCHA}, HyperGCN~\cite{HyperGCN}, HNHN~\cite{HNHN}, AllSet~\cite{Allset}, 
ED-HNN~\cite{wang2022equivariant},
SheafHGNN~\cite{duta2024sheaf}, and DPHGNN~\cite{saxena2024dphgnn}. 

\noindent\textbf{Experimental Settings.} We adopt accuracy to evaluate our model and baseline methods and split the data into training/validation/test samples using 20\%/20\%/60\% splitting percentages.  
Additionally, we conducted each method five times with 500 epochs and reported the average score with standard deviation. All experiments are conducted under the environment of the Ubuntu 22.04.3 OS plus an Intel i9-12900K CPU, two GeForce RTX 3090 Graphics Cards, and 64 GB of RAM. We utilize Adam~\cite{adam} as the optimizer. 
We run a grid search on hyper-parameters and report the best performance among the optimal hyper-parameters for each method across all datasets.
\begin{figure*}[t]
\begin{center}
    \includegraphics[width=0.85\textwidth]{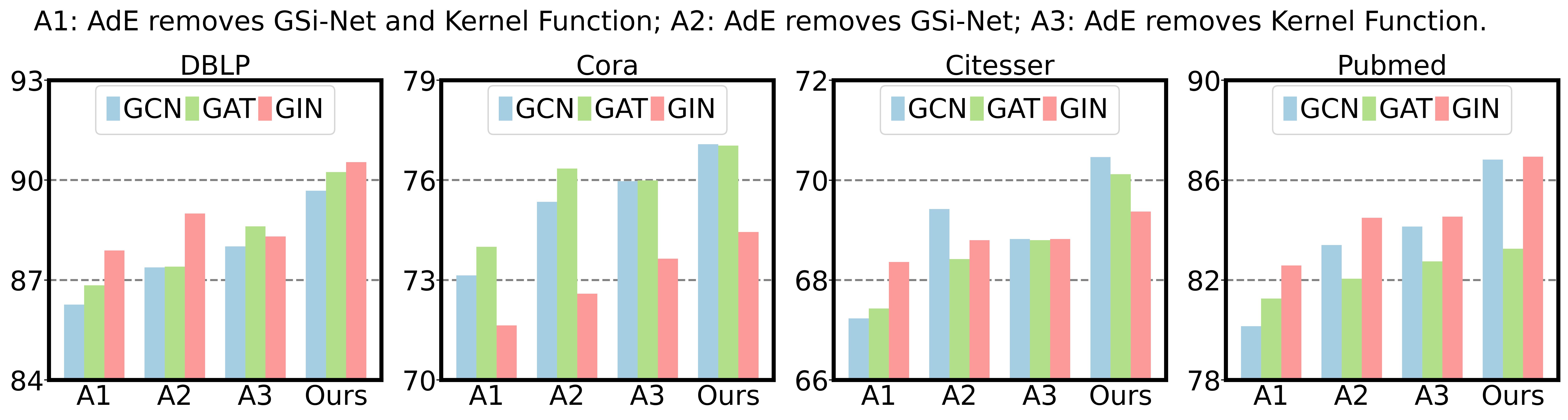}
\end{center}
\vspace{-4mm}
\caption{Performances of different model variants over DBLP, Cora, Citeseer, and Pumbed.}
\label{fig: ablation}
\vspace{-1mm}
\end{figure*}
\begin{figure*}[ht]
\begin{center}
    \vspace{-2mm}
    \includegraphics[width=0.8\textwidth]{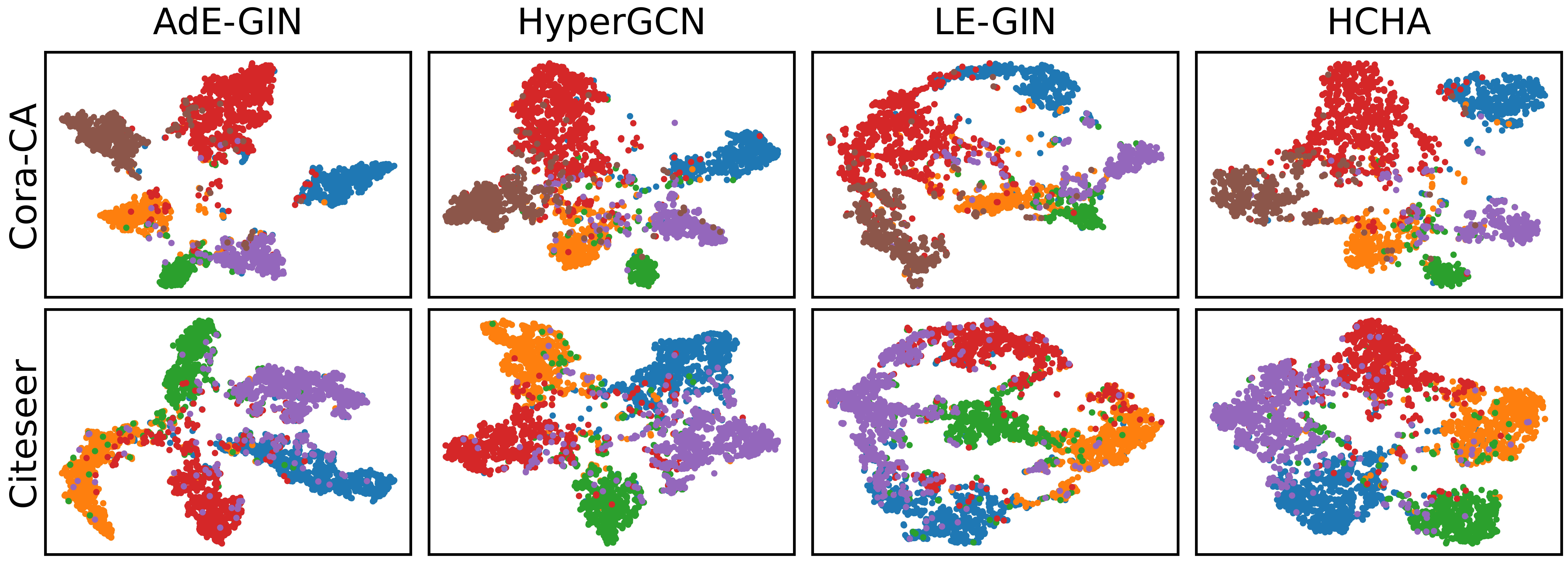}
\vspace{-2mm}
\caption{Embeddings visualization for AdE-GIN, HyperGCN, LE-GIN, and HCHA over two citation hypergraphs.}\label{fig: TSNE}
\end{center}
\vspace{-4mm}
\end{figure*}
\subsection{Performance Comparison} Table~\ref{table: combine performance 1 3} shows the accuracy performance among hypergraph expansions and our model over seven benchmark hypergraph datasets. 
According to Table~\ref{table: combine performance 1 3}, we draw the following conclusions: 
(i) Merely leveraging the classic CE method to expand hypergraphs is insufficient to depict complex high-order relationships among hypergraphs, as all models in G1 have the worst performance compared to other groups;
(ii) Other hypergraph expansion methods achieve relatively satisfactory performance over different hypergraphs, showing their advanced ability over some hypergraphs. 
For instance, SE-based methods show better performance over DBLP and Citeseer, Uni-GAT demonstrates the best performance among all expansion methods over Senate 0.6, Senate 1.0, and House 1.0, and LE-GAT is the best method over Cora-CA, Cora, and House 0.6.
(iii) Compared to all the methods, our AdE-based methods outperform all the baseline methods, which shows the strong effectiveness of AdE over both homophilic and heterophilic hypergraphs.
Table~\ref{table: combine performance 2 4} shows the accuracy performance among MLP, eight HyGNNs methods, and AdE on benchmark hypergraphs. 
Based on Table~\ref{table: combine performance 2 4}, we find that:
(i) All HyGNNs outperform the feature-based method MLP in homophilic hypergraphs, showing that the hypergraph structure enhances the performance for node classification tasks over homophilic hypergraphs to a large extent.
(ii) MLP outperforms some HyGNNs in heterophilic hypergraph datasets. This finding indicates that heterophily may have a negative effect on HyGNNs.
(iii) Our proposed method, AdE, outperforms all baseline methods in every hypergraph dataset,
which again shows that our model performs excellently on various hypergraphs.

\noindent\textbf{Ablation Study.} To show the effectiveness of each component in our framework, we conduct a set of ablation experiments over four benchmark hypergraph datasets for node classification tasks and further analyze the contribution of each component in our framework, i.e., our GSi-Net and Kernel function (A1), GSi-Net (A2),  kernel function (A3), by removing it separately, as illustrated in Fig.~\ref{fig: ablation}. 
First, we remove the GSi-Net and kernel function from our model, which means that we employ the CE method to obtain the graph and feed it to GNNs for representation learning.
We conclude that our adaptive expansion method is effective enough as
the performance of A1 drops significantly on all hypergraph datasets.
Afterward, we remove GSi-Net from our model, which means that we utilize the distance-aware kernel function to assign weights to the graph obtained via CE. 
The performance of A2 obviously decreases in all datasets, showing the effectiveness of the GSi-Net module. 
Moreover, we remove the kernel function from our model, which means that we merely employ the GSi-Net to obtain the graph while assigning the fixed edge weights. 
The decline in A3 shows that the kernel function has contributed to our model. 

\noindent\textbf{Embedding Visualization.} To further examine the effectiveness of our model intuitively, we render the embeddings of the Cora-CA and Citeseer datasets generated by AdE-GIN, HyperGCN, LE-GIN, and HCHA in Fig.~\ref{fig: TSNE}. 
Each unique color represents the embeddings corresponding to a specific class. 
According to Fig.~\ref{fig: TSNE}, our model shows more distinct boundaries and smaller overlapping areas compared to other baseline methods, again demonstrating the effectiveness of our expansion in learning hypergraph representation for node classification tasks. 

\noindent\textbf{Complexity Analysis.}
The time complexity to feed the node feature matrix into GSi-Net is linear to $\mathcal{O}(N)$, where $N$ is the number of nodes. 
Then, selecting hyperedge representative node pairs takes $\mathcal{O}(M)$ where $M$ is the number of hyperedges. The time complexity to compute the elements in distance matrix $\mathcal{U}$ and generate edges with weights takes $\mathcal{O}(E)$, where $E = \sum_{e\in \mathcal{E}} d(e)$. 
The total time complexity of AdE for each round is $\mathcal{O}(N + M + E) = \mathcal{O}(E)$. 

\section{Conclusion}
In this paper, we introduce a novel CE-based adaptive expansion method called AdE to address the limitations of existing hypergraph expansion methods.
We first introduce a novel global simulation network called GSi-Net to choose representative node pairs in an adaptive manner to symbolize every hyperedge.
Then we devise a distance-aware kernel function that dynamically adjusts the edge weights to ensure that nodes with similar attribute features within the corresponding hyperedge are more likely to be connected. 
Afterward, we leverage graph neural networks to model the complex interaction among nodes for downstream classification tasks.
We also provide extensive theoretical justifications and comprehensive experiments over seven benchmark hypergraphs, i.e., five homophilic datasets and two heterophilic hypergraphs, to demonstrate the effectiveness, rationality and generality of AdE, compared to classic expansion methods.

\bibliographystyle{named} 
\bibliography{main.bib}

\newpage
\appendix
\noindent{\Huge Appendix}
\section{Algorithm}\label{appendix: algorithm}
\begin{algorithm}[htbp]
    \caption{Training Procedure of AdE}\label{alg}
    \begin{algorithmic}
        \STATE{\bfseries Input:} Hypergraph $\mathcal{H}$, GSi-Net, distance-aware kernel function, and GNN encoder $f(\cdot)$.
        \STATE {\bfseries Initialize:} The distance matrix $\mathcal{U}$. 
        \FOR{each epoch $t$}
        \STATE Feed $\mathcal{H}$ into GSi-Net to learn the scaled feature $\rmX_\text{a}$, and further select representative nodes $(v_{e^-}, v_{e^+})$ based on $\rmX_\text{a}$.
        \STATE Compute the distance-aware edge weight $\mathcal{W}$ via Eq.~\ref{eq: weight func}.
        \STATE Normalize the edge weight via Eq.~\ref{eq: normalized weight function}.
        \STATE Obtain the adaptive weighted graph $\mathcal{G}_\text{a} = (\mathcal{V},\mathcal{E}_\text{a}, \mathcal{X}_\text{a})$.
        \STATE Feed weighted graph $\mathcal{G}_\text{a}$ into GNN encoder $f(\cdot)$ to generate node embedding $\rmZ$ via Eq.~\ref{eq: nodeembed}.
        \STATE Optimize the GNN encoder $f(\cdot)$, the parameters in  AdE by minimizing the cross-entropy loss $\mathcal{L}$.
        \ENDFOR
    \end{algorithmic}
\end{algorithm}
\section{Data Description} \label{appendix: data}
To evaluate the effectiveness of our model, we employ five benchmark homophilic hypergraph datasets adapted
from~\cite{HyperGCN}: the coauthorship networks, i.e., Cora-CA and DBLP; the cocitation
networks, i.e., Cora, Citeseer, and Pubmed. In both coauthor hypergraph datasets, documents co-
authored by an author are connected via one hyperedge. In three cocitation hypergraph datasets, all
documents referenced by a document are connected by a hyperedge. Moreover, we adopt two benchmark heterophilic hypergraphs, i.e., Senate and House~\cite{wang2022equivariant,veldt2021higher}.
In the Senate dataset, nodes
are members of the US Senate, and hyperedges correspond to committee memberships. 
In the House hypergraph, nodes are
members of the US House of Representatives, and hyperedges correspond to committee memberships. 
Since the Senate and House datasets do not contain node features, we leverage label-dependent Gaussian distribution~\cite{deshpande2018contextual} to create node features.
Table~\ref{table: data stat} lists the statistics of
seven benchmark hypergraph datasets.
\begin{table*}[htbp]
\vspace{-1mm}
    \caption{The statistics of benchmark hypergraph datasets.}
    \vspace{-1mm}
    \label{table: data stat}
    \centering
    \setlength\tabcolsep{12pt}
    \begin{tabular}{l c c c c c c c}
    \toprule
    & Cora-CA & DBLP & Cora & Citeseer & Pubmed & House & Senate    \\
    \midrule
      \# nodes   & 2,708 & 41,302 & 2,708 &  3,312 & 19,717 & 1,290 & 282  \\ 
      \# hyperedges  & 1,072  & 22,363 & 1,579 & 1,079 & 7,963 
 & 340 & 315 \\
      \# features  & 1,433 & 1,425 & 1,433 & 3,703 & 500 & 100 & 100  \\
      \# class & 7 & 6 & 7 & 6 & 3 & 2 & 2 \\
        avg. $d(e)$  & 4.28 & 4.45 & 3.03 & 3.20 & 4.35 & 34.73 & 17.17 \\
        avg. $d(v)$  & 1.69 &  2.41 & 1.77 & 1.04 & 1.76 & 9.18 & 19.18  \\
      \bottomrule
    \end{tabular}
    \vspace{-4mm}
\end{table*}
\section{Baseline Settings}~\label{appendix: baseline}
We compare our model with MLP and seven HyGNNs, including HGNN~\cite{HGNN}, HyperGCN~\cite{HyperGCN}, HNHN~\cite{HNHN}, HCHA~\cite{HCHA}, UniGCNII~\cite{UniGNN}, ALlSet~\cite{Allset}, ED-HNN~\cite{wang2022equivariant}, SheafHyGNN~\cite{duta2024sheaf}, and DPHGNN~\cite{saxena2024dphgnn}.
MLP primarily focused on encoding the node attribute features while ignoring the graph structures. We adopt a two-layer MLP on attribute features to generate node representations for classification.
HGNN employs the convolution operation using truncated Chebyshev polynomials to generate node representations. 
HyperGCN leverages hypergraph Laplacian to convert hypergraphs into weighted graphs and feeds the weighted graph into GCN to learn node embeddings.
HCHA introduces an attention mechanism to learn the importance of nodes in hypergraphs and utilizes neural networks to generate node embeddings. 
HNHN extends HGNN with nonlinear activation functions applied to both nodes and hyperedges, combined with a normalization scheme that adjusts the importance of high-cardinality hyperedges and high-degree nodes. 
UniGCNII leverages the pooling function to learn the hyperedge representation through node attribute features and utilizes GNNs to propagate hyperedge representations back to study the node representations. 
AllSet unifies a whole class of two-stage message-passing models with multiset functions.
ED-HNN employs equivariant operators to model the hypergraph diffusion process.
SheafHGNN introduces neural sheaf diffusion models that learn cellular
sheaves in hypergraphs.
DPHGNN employs a dual-perspective HyGNN that introduces an equivariant operator to capture lower-order semantics by inducing topology awareness of spatial and spectral inductive biases.
We strictly follow the settings of baseline methods from their source code to reproduce the experimental results. 

\section{Proofs}\label{appendix: proof}

\subsection{Proof of Proposition~\ref{proposition: w}}\label{proof: w}
First of all, for a node pair $(v_i,v_j)$, we calculate the distance among each dimension to quantify the closeness of the node pair and employ the pre-computed distance matrix in the original feature space as the prior information to guide the learning process of the edge weight.
    Formally, we have: 
    \vspace{-1.5mm}
    \begin{equation}
        \vspace{-1mm}
        \Delta_{d,(i,j)}^2 = \gU_{i,j}(\mathbf{X}_{\text{a},({i,d})} - \mathbf{X}_{\text{a}, (j, d)})^2. 
    \end{equation}
    Afterward, we introduce a learnable parameter $\theta\in\mathbb{R}^{b}$ to control the sensitivity of the node feature similarity across $b$ dimensions, while a larger $\theta_d$ would make it less sensitive, and vice versa.
    The average of values $\Delta_{d, (i,j)}^2$ across $b$ dimensions is computed as follows:
    \vspace{-3mm}
    \begin{equation}
        \Gamma_{i,j} = \frac{1}{b}\sum\nolimits_{d=1}^b \frac{\Delta_{d,(i,j)}^2}{\theta_d^2}.
    \end{equation}
    Finally, we compute the edge weight by feeding the negative $\Gamma_{i,j}$ into an exponential function $\gW_{i,j} = \exp(-\Gamma_{i,j}).$ 
    
    Considering two nodes occupy dissimilar behaviors, the distance in $\Delta_{d,(i,j)}^2$
    is supposed to be larger, and $\theta_d$ learns to facilitate the similarity measures. 
    In this case, the negative value $-\Gamma_{i,j}$ will be smaller, resulting in a smaller edge weight value $\mathcal{W}_{i,j}$. 
    Therefore, two dissimilar nodes would have a smaller edge weight and vice versa. This completes the proof.
\subsection{Proof of Proposition~\ref{proposition: 3}} \label{proof: 3}
Given the same selected nodes $(v_{e^-}, v_{e^+})$ for hyperedge $e$, HyperGCN fixes the edge weights for all node pairs within the same hyperedge.
Specifically, as the size of edges $\gE_e$  is $2|e| -3$ for hyperedge $e$ in the converted graph, the weight for any edge in $\gE_e$ is $(2|e| -3)^{-1}$.

In the worst case, the kernel function $\mathcal{W}$ in Eq.~\ref{eq: weight func} assigns the edge weight $\mathcal{W}_{i,j}$ of any edge $(v_i, v_j) \in \gE_e$ remains the same, i.e., an arbitrary positive values $\epsilon^{(e)}\in \sR^+$. Then we obtain the normalized edge weight in Eq.~\ref{eq: normalized weight function} as follows:
\vspace{-2mm}
\begin{equation}
    \Bar{\gW}_{i,j}^{(e)} = \frac{\epsilon^{(e)}}{\sum_{\{v_k,v_g\} \in \gE_e}\epsilon^{(e)}} = \frac{1}{2|e|-3},
    \vspace{-1mm}
\end{equation}
which is the same as HyperGCN.
However, according to Proposition~\ref{proposition: w}, the kernel function $\mathcal{W}$ in Eq.~\ref{eq: weight func} assigns higher edge weights for node pairs with similar features, and vice versa. We have $\mathcal{W}_{i,j} > \epsilon^{(e)}$, and $\Bar{\mathcal{W}}_{i,j}^{(e)}  > (2|e| -3)^{-1}$. 
Therefore, we conclude that AdE enhances HyperGCN in generating more adaptive edge weights when they select identical representative node pairs $(v_{e^-}, v_{e^+})$. This completes the proof. 

\subsection{Proof of Proposition~\ref{proposition: CE}} \label{proof: CE}
    Given a hypergraph $\mathcal{H}=(\mathcal{V}, \mathcal{E}, \mathcal{X})$, as mentioned by~\cite{CE}, the weighted graph $\mathcal{G}_\text{c}$ via CE can be represented by the following adjacency matrix:
    \begin{equation}\label{eq: WCE-adj}
         A_{\text{c},(i, j)} = \sum\nolimits_{e\in\mathcal{E}} \rmH_{i,e}\rmH_{j,e} w_{i,j},
    \end{equation}
    where $\rmH$ denotes the incident matrix, and $w$ is an edge weight function.
    Here, we use $ \Bar{\mathcal{W}}_{i,j}^{(e)}$ in Eq.~\ref{eq: normalized weight function} as the criteria to compute edge weights. 
    Afterward, the adjacency matrix generated by AdE is formulated as:
    \begin{equation}
        A_{\text{a},(i,j)} = \sum\nolimits_{e\in\mathcal{E}} \mathbb{I}\,[\{v_i, v_j\}\in\mathcal{E}_e]\Bar{\mathcal{W}}_{i,j}^{(e)}.
    \end{equation}
    Consider a 3-uniform hypergraph where each hyperedge contains three nodes. 
    AdE chooses two nodes as $v_{e^+}$ and $v_{e^-}$, with the remaining one as the mediator. Edges are then generated to connect any two nodes within the corresponding hyperedge.

    Since there is only one mediator for each hyperedge in a 3-uniform hypergraph, 

    AdE will produce a graph structure identical to that generated by the CE.
    As the weight matrix $w$ in Eq.~\ref{eq: WCE-adj} can be any form, we can replace $w$ with $\Bar{\mathcal{W}}$. 
    Therefore, AdE is equivalent to the weighted clique expansion in a 3-uniform hypergraph. This completes the proof.
\subsection{Proof of Lemma~\ref{lemma: mapping}}\label{proof: mapping}

    We prove this by contradiction. Assume at iteration $t>0$, for node $v_i\in\gV_\gH$ and $v_j\in\gV_{\gH'}$, AdE constructs edges $E_{v_i}$ for graph $\gG$ and $E_{v_j}'$ for graph $\gG'$:
    \begin{align}
        E_{v_i} &= \Big\{ \{ v_i, v_k \} \Big| v_k\in e, e\in \gE_{v_i}, v_i \oplus v_k\in\gV_\text{m}^e  \Big\}, \\
        E_{v_j}' &= \Big\{ \{ v_j, v_k \} \Big| v_k\in e', e'\in \gE_{v_j}', v_j \oplus v_k\in\gV_\text{m}^{e'}  \Big\},  
    \end{align}
    such that edge sets $E_{v_i}$ and $E_{v_j}'$ cannot map to each other.
     Here, $E_{v}^*$ denotes the edges incident to node $v$ in graph $\gG^*$, $\gE_{v}^*$ is the hyperedges that contain node $v$ in hypergraph $\gH^*$, $\gV_\text{m}^{e*}$ represents the node mediators of hyperedge $e$ discussed in Section~\ref{section: gsi-net}, and $\oplus$ is the XOR operation.  
    Based on the assumption, there is no bijective function $\pi$ maps nodes in $\gV_\gG$ to nodes in $\gV_{\gG'}$  so that the edge set $E_{v_i}$ is equivalent to edge set $E_{v_j}'$. Mathematically, we assume:
    \vspace{-1mm}
        \begin{align}
            \nexists \pi: \gV_\gG \rightarrow\gV_{\gG'}, \text{s.t.} \biggl\{\Bigl\{\pi(v_i), \pi(v_k)\Bigr\} \bigg| \{v_i, v_k\} \in E_{v_i}\biggr\} = E_{v_j}' .
            \vspace{-1mm}
        \end{align}  
    However, as 1-GWL assigns the same label for node $v_i\in\gV_\gH$ and node $v_j\in\gV_{\gH'}$ at iteration $t > 0$, there is an bijective function $\psi: \gE_{v_i}\rightarrow\gE_{v_j}'$ that maps each hyperedge $e\in\gE_{v_i}$ to hyperedge $e'\in\gE_{v_j}'$,
    and it is also applicable to labels for hyperedges, i.e., $\forall e \in \gE_{v_i}, h_{\psi(e)}^{(t)} = h_{e'}^{(t)}, \psi(e) = e'\in\gE_{v_j}'$. Moreover, we can find another mapping function $\rho: \gV_\gH\rightarrow \gV_{\gH'}$ for nodes between $\gV_\gH$ and $\gV_{\gH'}$ upon the mapping $\psi$:
    \begin{align}
        &\forall v_i\in\gV_\gH, \rho(v_i) = v_j\in\gV_{\gH'}, \nonumber\\
        &\text{ s.t. } h_{v_i}^{(t)} \in h_e^{(t)},  h_{v_j}^{(t)} \in h_{\psi(e)}^{(t)}, h_{v_i}^{(t)} =    h_{v_j}^{(t)}.
    \end{align}
    As AdE can be viewed as a bijective function for nodes between the hypergraph and the converted graph, there is a contradiction with our assumption. This completes the proof.

\subsection{Proof of Theorem~\ref{theorem 1GWL 1WL}} \label{proof: 1GWL 1WL}
    At iteration $t=0$, it is trivial to prove as all node labels in graphs $\gG$ and $\gG'$, all node and hyperedge labels in hypergraph $\gH$ and $\gH'$ are initialized in the same way. 
    Assume at iteration $t \in (0, T)$, 1-GWL test and 1-WL test obtain:
    \begin{align} 
        \label{eq: 1-GWL equal}
        \{\{h_{v_i}^{(t)}|v_i\in \gV_\gH\}\} &= \{\{h_{v_j}'^{(t)}|v_j\in \gV_{\gH'}'\}\},    \\ 
        \label{eq: 1-WL equal} 
        \{\{l_{v_p}^{(t)}|v_p\in \gV_\gG\}\} &= \{\{l_{v_q}'^{(t)}|v_q\in \gV_{\gG'}'\}\}, \text{respectively}. 
    \end{align} 
    Here, $\gV^*_{\gH^*}$ is the node set of hypergraph $\gH^*$, and $\gV^*_{\gG^*}$ is the node set of graph $\gG^*$. Eq.~\ref{eq: 1-GWL equal} and ~\ref{eq: 1-WL equal} demonstrate bijective mappings $\pi_*^{(t)}: \gV \rightarrow \gV'$, s.t., $  \forall v\in\gV,  \pi_*^{(t)}(v) = u\in\gV'$, where $*$ is either $\gH$ or $\gG$, representing the bijective mapping $\pi_\gH$ for hypergraphs or $\pi_\gG$ for graphs.
    These mappings are also applicable to labels:
    \vspace{-2mm}
    \begin{align}
        &\forall v_i\in \gV_\gH, h_{v_i}^{(t)} =  h_{\pi_{\gH}(v_i)}'^{(t)} = h_{v_j}'^{(t)}, v_j\in\gV_{\gH'}', \\
        &\forall v_p\in \gV_\gG, l_{v_p}^{(t)} = l_{\pi_{\gG}(v_p)}'^{(t)} = l_{v_q}'^{(t)}, v_q\in\gV_{\gG'}'.     
    \vspace{-2mm}
    \end{align}
    At iteration $t+1$, 1-GWL test fails to distinguish hypergraphs $\gH$ and $\gH'$. We also have:
    \begin{align}
        \{\{h_{v_i}^{(t+1)}|v_i\in \gV_\gH\}\} & = \{\{h_{v_j}'^{(t+1)}|v_j\in \gV_{\gH'}'\}\}, \text{ and}\\
        \{\{h_{e_i}^{(t+1)}|e_i\in \gE_\gH\}\} & = \{\{h_{e_j}'^{(t+1)}|e_j\in \gE_{\gH'}'\}\}. 
    \end{align}
    Notice that the mapping $\pi_{\gH}^{(t)}$ still holds at iteration $t+1$ for nodes in hypergraphs $\gH$ and $\gH'$:
        \begin{align}
            \forall v_i\in \gV_\gH, h_{v_i}^{(t+1)} = h_{\pi_{\gH}(v_i)}'^{(t+1)} = h_{v_j}'^{(t+1)}, v_j\in\gV_{\gH'}'. \label{eq: 1GWL 1WL edge}
        \end{align}
    According to Lemma~\ref{lemma: mapping} and Eq.~\ref{eq: 1GWL 1WL edge}, we have a bijective function $\pi_{v_i}$ for each node $v_i\in\gV_\gG$:
    \begin{align} \label{eq: theorem1 mapping for graph nodes}
        &\exists \pi_{v_i}: \gV_\gG \rightarrow\gV_{\gG'}',\nonumber\\
        &\text{s.t.} \biggl\{\Bigl\{\pi(v_i), \pi(v_k)\Bigr\} \bigg| \{v_i, v_k\} \in E_{v_i}\biggr\} = E_{v_j}'.
    \end{align}
    The labels of nodes $v_i\in\gV_\gG$ and $v_j\in\gV_{\gG'}'$ are updated as:
        \begin{align}
           l_{v_i}^{(t+1)} &=\Big\{\Big\{ (l_{v_i}^{(t)}, l_{v_p}^{(t)}) \Big\}\Big\}_{v_p\in\gN_i},\\
           l_{v_j}'^{(t+1)} &=\Big\{\Big\{ (l_{v_j}'^{(t)}, l_{v_q}'^{(t)}) \Big\}\Big\}_{v_q\in\gN_j}. \label{eq: theorem1 graph G' label 1}
        \end{align}
    Notice that $\gN_i$ and $\gN_j$ denote the neighbors of node $v_i$ and node $v_j$, respectively. 
    According to Eq.~\ref{eq: theorem1 mapping for graph nodes}, the node $v_i\in\gV_\gG$ and its neighbors can mapping to node $v_j\in\gV_{\gG'}'$ and its neighbors via bijective mapping $\pi_{v_i}$, i.e., $\forall v_p\in\gN_i \cup \{v_i\}, \pi_{v_i}(v_p) = v_q \in\gN_j \cup \{v_j\}$. Moreover, the Eq.~\ref{eq: theorem1 graph G' label 1} can be rewrite in terms of nodes in $\gN_i$:
    \vspace{-1mm}
    \begin{align}
        l_{v_j}'^{(t+1)} &=\Big\{\Big\{ (l_{\pi_{v_i}(v_i)}'^{(t)}, l_{\pi_{v_i}(v_p)}'^{(t)}) \Big\}\Big\}_{v_p\in\gN_i} \nonumber \\ 
        & = \Big\{\Big\{ (l_{v_i}^{(t)}, l_{v_p}^{(t)}) \Big\}\Big\}_{v_p\in\gN_i} =  l_{v_i}^{(t+1)}.
        \vspace{-1mm}
    \end{align}
    Here, we show that, at iteration $t+1$, for each node $v_i\in\gV_\gG$, the labels of nodes $v_i$ and $v_j=\pi_{v_i}(v_i)\in\gV_{\gG'}'$ are identical, which means 1-WL cannot differentiate the converted graphs $\gG$ and $\gG'$. This completes the proof.
\subsection{Proof of Theorem~\ref{theorem 1WL 1GWL}} \label{proof: 1WL 1GWL}
Assume 1-WL test decides converted graphs $\gG=\text{AdE}$ and $\gG' = \text{AdE}(\gH')$ are non-isomorphic at time $t>0$, then by the Definition~\ref{def: 1-WL}, in the worst case,  we have:
\begin{align}
    &\{\{l_{v_i}^{(t)} | v_i\in\gV_\gG \}\} =  \{\{l_{v_j}^{(t)} | v_j\in\gV'_{\gG'} \}\}, \nonumber\\
    &\exists v_p\in\gV_\gG, v_q\in\gV'_{\gG'}, \text{ s.t. }  l_{v_p}^{(t+1)}  \ne l_{v_q}^{(t+1)}, \text{and }  \label{eq: theorem2 assumption1} \\ 
     &\{\{l_{v_i}^{(t+1)} | v_i\in\gV_\gG/ v_p \}\} =  \{\{l_{v_j}^{(t+1)} | v_j\in\gV'_{\gG'}/v_q \}\}. \nonumber
\end{align}
1-WL updates the label of node $v_i\in\gV_\gG$ as follows:
\begin{align}\label{eq: theorem2 1WL update rule}
    l_{v_i}^{(t+1)} = \left\{\left\{ (l_{v_i}^{(t)}, l_{v_k}^{(t)}) \right\}\right\}_{v_k\in\gN_i}.
\end{align}
According to Eq.~\ref{eq: theorem2 1WL update rule} and Eq.~\ref{eq: theorem2 assumption1}, we have:
\begin{align} \label{eq: theorem2 1WL node mapping}
    \nexists \psi: \gV_\gG \rightarrow \gV_{\gG}', \text{ s.t. } \left\{ \psi(v_k) | v_k\in\gN_p \right\} = \gN_q. 
\end{align} 
As AdE converts hypergraphs with identical features and structures into the same graph, from Eq.~\ref{eq: theorem2 assumption1} and Eq.~\ref{eq: theorem2 1WL node mapping}, we conclude that:
\begin{align}
    &\exists e_p\in\gE_\gH \land e_q\in\gE'_{\gH'}, \text{ s.t. }
      \forall \rho: \gE_\gH \rightarrow \gE'_{\gH'} h_{\rho(e_p)}^{(t+1)}  \ne h_{e_q}^{(t+1)} , \nonumber \\ 
    &\{\{ h_{\rho(e)}^{(t+1)} | e\in \gE_\gH / e_p \}\} =  \{\{ h_{e'}^{(t+1)} | e'\in \gE'_{\gH'} / e_q \}\}. 
\end{align}
Therefore, the hyperedge labels in hypergraphs $\gG_\gH$ and $\gG'_{\gH'}$ are distinct, i.e., $\{\{ h_{e}^{(t+1)} | e\in \gE_\gH \}\} \ne  \{\{ h_{e'}^{(t+1)} | e'\in \gE'_{\gH'} \}\} $, and this leads to the different node labels for hypergraphs: $\{\{ h_{v}^{(t+1)} | v\in \gV_\gH \}\} \ne  \{\{ h_{v'}^{(t+1)} | v'\in \gV'_{\gH'} \}\}$. The 1-GWL test decides $\gH \ne \gH'$ at iteration $t$. This completes the proof.

\end{document}